%% file: neurips_data_2024.tex
\documentclass{article}

\usepackage[preprint,nonatbib]{neurips_data_2024}

\usepackage[utf8]{inputenc} 
\usepackage[T1]{fontenc}    
\usepackage{url}            
\usepackage{booktabs}       
\usepackage{amsfonts}       
\usepackage{nicefrac}       
\usepackage{microtype}      
\usepackage{xcolor}         

\usepackage[pagebackref=true,breaklinks=true,letterpaper=true,colorlinks,bookmarks=false]{hyperref}

\usepackage{multirow}
\usepackage{adjustbox}
\usepackage{booktabs}
\usepackage{amsmath}
\usepackage{enumitem}
\usepackage{wrapfig}
\usepackage{graphicx}
\usepackage{subfig}

\title{GenBench: A Benchmarking Suite for Systematic Evaluation of Genomic Foundation Models}

\author{Zicheng Liu$^{1,2,}$\thanks{Equal contribution.\quad $^\dag$Stan Z. Li (stan.zq.li@westlake.edu.cn) is the corresponding author.}
\quad\quad Jiahui Li$^{1,*}$\quad\quad Siyuan Li$^{1,2}$\quad\quad Zelin Zhang$^{1,2}$\quad\quad \vspace{0.2em}\\
\textbf{Cheng Tan}$^{1,2}$\quad\quad\textbf{Yufei Huang$^{1,2}$\quad\quad Yajing Bai$^{3}$\quad\quad Stan Z. Li$^{1,\dag}$}\\
AI Lab, Research Center for Industries of the Future, Hangzhou, China; \\
$^1$Westlake University;\quad$^2$Zhejiang University; \quad$^3$University of Southern California; \\
\vspace{0.2em}
\texttt{\{liuzicheng; lijiahui\}}\texttt{@westlake.edu.cn}
}

\begin{document}

\maketitle

\begin{abstract}
  The Genomic Foundation Model (GFM) paradigm is expected to facilitate the extraction of generalizable representations from massive genomic data, thereby enabling their application across a spectrum of downstream applications. 
  Despite advancements, a lack of evaluation framework makes it difficult to ensure equitable assessment due to experimental settings, model intricacy, benchmark datasets, and reproducibility challenges.
  In the absence of standardization, comparative analyses risk becoming biased and unreliable. To surmount this impasse, we introduce GenBench, a comprehensive benchmarking suite specifically tailored for evaluating the efficacy of Genomic Foundation Models. GenBench offers a modular and expandable framework that encapsulates a variety of state-of-the-art methodologies. Through systematic evaluations of datasets spanning diverse biological domains with a particular emphasis on both short-range and long-range genomic tasks, firstly including the three most important DNA tasks covering Coding Region, Non-Coding Region, Genome Structure, etc. Moreover, We provide a nuanced analysis of the interplay between model architecture and dataset characteristics on task-specific performance. Our findings reveal an interesting observation: independent of the number of parameters, the discernible difference in preference between the attention-based and convolution-based models on short- and long-range tasks may provide insights into the future design of GFM. 
  We made the GenBench codebase and associated models publicly accessible \href{https://github.com/jimmylihui/OpenGenome}{here}.
\end{abstract}

\input{1_intro}
\input{2_background}
\input{3_method}
\input{4_exps}
\input{5_conclusions}

\bibliographystyle{unsrt}
\bibliography{reference}
\input{7_checklist}

\newpage
\input{6_appendix}

\end{document}

%% file: 1_intro.tex
\section{Introduction}

In recent years, significant advancements have been made in genomic research by utilizing foundation models (FMs) to analyze unstructured whole genome data. These genomic foundation models play a crucial role in various tasks such as predicting gene locations and functions, identifying regulatory elements, and studying species evolution~\cite{ji2021dnabert,fishman2023gena,zvyagin2023genslms}.

Despite the importance of modeling Genomics foundations and the advancement of different training methods, there is still a noticeable absence of a thorough benchmark in this area that encompasses a range of practical application scenarios and different foundational model structures. The current benchmark either restricts its scope to short distances or oversimplifies the challenge by focusing solely on the classification task~\cite{ji2021dnabert,marin2023bend,fishman2023gena}. Moreover, with the influx of long sequence models~\cite{gu2023mamba,liu2024longvq,nguyen2024hyenadna,schiff2024caduceus}, a systematic approach to evaluating up-to-date GFMs and inspiring subsequent development is also sorely lacking.
Based on the current state of research, we thus summarise the three key problems in current GFMs:
\textbf{(1) Incomplete evaluation:} Long sequence processing is crucial for modeling genetic data. Current tests of these models for long-sequence gene tasks are incomplete. 
\textbf{(2) Chaotic training strategies:}  The variety of tokenizations and pre-training methods lack a fair platform to compare and select the most appropriate training method for DNA data. 
\textbf{(3) Snowy model design:} Do the various Attention-based and recent state-space models each have strengths in DNA? We need an experience that inspires subsequent designs.

\begin{figure*}[t!]
  \centering
  \includegraphics[width=0.9\textwidth]{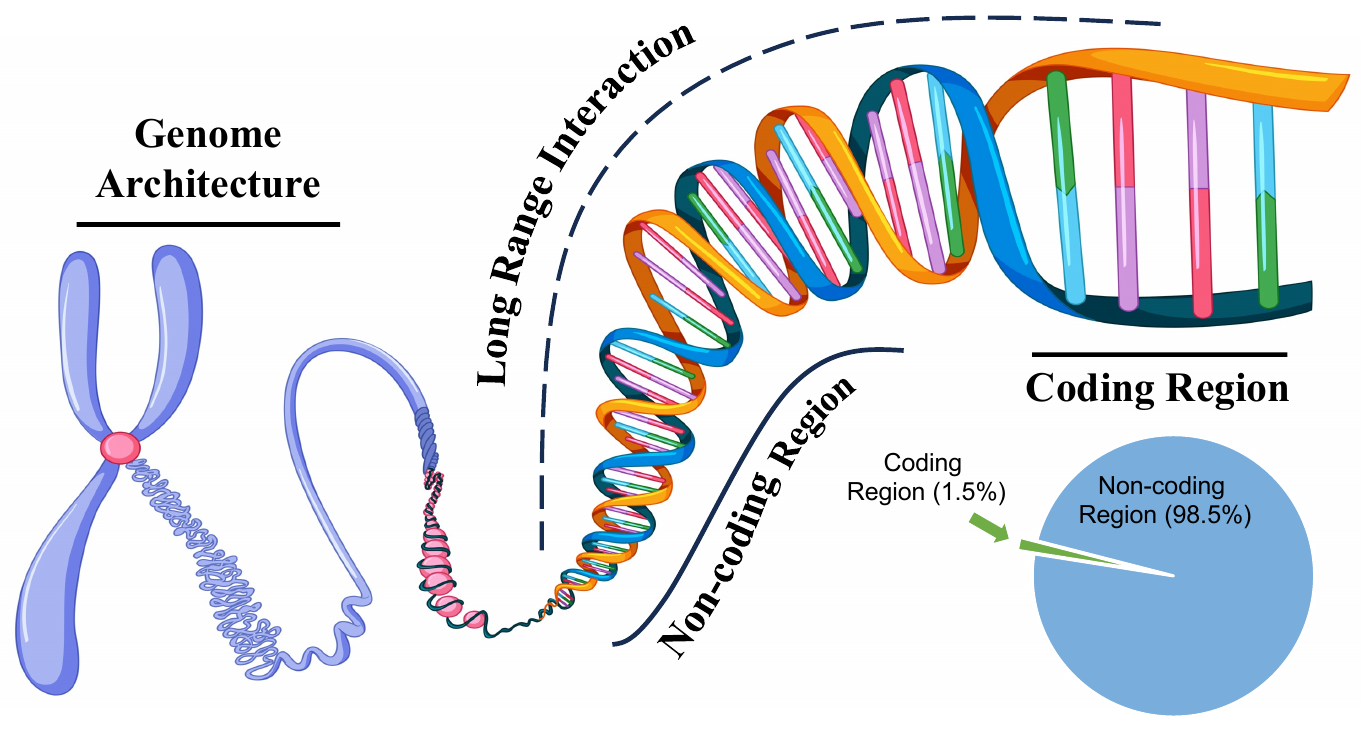}
  \caption{The image illustrates the intricate organization of eukaryotic genomic DNA, highlighting three critical components: the Coding Region, the Non-coding Region, and Genome Architecture. The Coding Region, comprising only about 1.5\% of the genome, is responsible for coding proteins, essential for cellular functions. In contrast, the vast Non-coding Region, making up 98.5\% of the genome, plays crucial roles in gene regulation and genome stability, containing regulatory elements like promoters, enhancers, and silencers, as well as structural components. Genome Architecture is depicted by the DNA being tightly packed around histone proteins into chromosomes, ensuring efficient storage and regulation of genetic information. Notably, long-range interactions within the DNA are essential for life processes, facilitating the regulation of gene expression by bringing distant regulatory elements into proximity with target genes.}
  \label{intro:overview}
  \vspace{-1em}
\end{figure*}

To address these issues, we propose GenBench, a comprehensive benchmarking suite covering the three main genomic directions from local to global, as shown in Figure~\ref{intro:overview}: the coding regions, the non-coding regions, and the genome architecture. GenBench aims to provide a standardized platform that measures the ability of GFMs to capture the intricacies of genomic data and to help advance this nascent field. 
Specifically, GenBench encompasses ten popular GFMs and performs extensive experiments across forty-three realistic datasets.
Our contribution is summarized as follows:
\begin{itemize}[leftmargin=*]
    \item \textbf{Comprehensive forty-three datasets:} We first integrate short- and long-range tasks covering three main aspects of genomics: non-coding region, coding region, and genome architecture.
    
    \item \textbf{Experiments covering various types of GFMs:} Investigate the impact of employing attention and convolution models in Genomic modeling on different scales.
    
    \item \textbf{A modular and expandable code framework:} Provide a unified experimental environment to achieve fair comparisons and facilitate subsequent development of new methods.

    \item \textbf{New results:} Independent of the number of parameters, GFMs based on convolution-based and attention-based structures have their strengths in downstream tasks. The nature of DNA data is closer to natural language for short-range tasks and to images for long-range tasks.
\end{itemize}

%% file: 2_background.tex
\section{Background and Related Work}
\subsection{Problem Definition}

\paragraph{Pre-training} We present the formal definition of the genomic modeling problem as outlined below. During the pretraining phase, the input $\mathcal{X}^{l, L}=\left\{\boldsymbol{x}^i\right\}_{l-L+1}^l\in\{A,G,C,T,N\}$ up to position $l$ covers the past $L$ frames of base pairs, comprising Adenine (A), thymine (T), cytosine (C), guanine (G), or not known (N) nuclotides. At this stage, specific portions of nucleotides are masked for prediction purposes, either predicting the masked nucleotides or the subsequent nucleotide $\mathcal{Y}^{l^{\prime}, L^{\prime}}=\left\{\boldsymbol{x}^i\right\}_{l^{\prime}-L^{\prime}+1}^{l^{\prime}}$ from position $l^{\prime}$. The genomic sequence is initially encoded by a tokenizer into tokenization tensors $\mathcal{Z}^{l,L,D}\in\mathbb{Z}^{L\times D}$ with a hidden dimension of D.

\paragraph{Fine-tuning} The model with learnable parameters $\Theta$ establishes a mapping $\mathcal{F}_{\Theta}: \mathcal{Z}^{l,L,D} \mapsto \mathcal{Y}^{l^{\prime}, L^{\prime}}$ by ultilizing nucleotide dependencies. In this context, $\mathcal{F}_{\Theta}$ represents a neural network trained to minimize the difference between the predicted target nucleotide and the actual nucleotide. The optimal parameters $\Theta^*$ are determined as:
\begin{equation}
    \Theta^*=\arg \min _{\Theta} \mathcal{L}\left(\mathcal{F}_{\Theta}\left(\mathcal{Z}^{l, L, D}\right), \mathcal{Y}^{l^{\prime}, L^{\prime}}\right)
\end{equation}
where $\mathcal{L}$ is a loss function that measures this disparity.
In this research, we classify prevalent downstream tasks into two categories: classification and regression. For the classification task, the target prediction is the classification of the input genomic sequence $\mathcal{Y}^{D^{\prime}}\in \{0,1\}^D$ , where D represents the number of classes. In regression tasks, target prediction is the numerical tensor  $\mathcal{Y}^{L^{\prime},D^{\prime}}\in \mathbb{R}^{L^{\prime}\times D^{\prime}}$, with $L^{\prime}$ indicating the position and $D^{\prime}$ representing the property dimension.

\subsection{Genomic Foundation Models}

\paragraph{Attention-based} Foundation models in deep learning are trained on extensive data sets using self-supervised learning. The importance of these models has grown due to their capacity to leverage large amounts of unlabeled data. For instance, DNABERT by \cite{ji2021dnabert} was developed based on the BERT model \cite{devlin2018bert} with a k-mer genomic tokenizer. Additionally, \cite{benegas2023dna} introduced the Genomic Pre-trained Network (GPN) for predicting non-coding variant effects, surpassing supervised learning methods. Researchers have explored different approaches to enhance performance. For example, \cite{dalla2023nucleotide} introduced NT (Nucleotide Transformer), a genomic model with billions of parameters. On the other hand, researchers focus on optimizing model components. \cite{zhou2023dnabert} proposed DNABERT-2, which replaces k-mer tokenization with Byte Pair Encoding (BPE) \cite{sennrich2015neural}. 

\paragraph{Convolution-based} Despite the computational cost associated with scaling up in sequence length due to the quadratic complexity of attention mechanisms, there is room for more efficient alternatives. HyenaDNA\cite{nguyen2024hyenadna}  and Caduceus utilize the hyena operator \cite{poli2023hyena} and state space model \cite{gu2023mamba} with a complexity of $\mathcal{O}(L\log_2L)$ and $\mathcal{O}(L)$, significantly lower than the $\mathcal{O}(L^2)$ of attention-based models.

%% file: 3_method.tex
\section{GenBench}
\subsection{Overview}

\begin{table}[t!]
    \vspace{-1.0em}
\caption{Classification of the supported Genomic foundation models in GenBench. The use of BERT in training strategy refers to training the model to predict the masked token in a sequence, while employing GPT in training strategy involves utilizing the next token prediction. Additionally, we have included expert models for particular downstream tasks that utilize One-hot encoding and training from the beginning for comparison.}

\begin{adjustbox}{width=0.9\columnwidth,center}

\begin{tabular}{ccccc}
\toprule
Category                         & Model                  & Conference/Journal  & Tokenizer & Training strategy \\ \hline
\multirow{4}{*}{Attention-based} & DNABERT                & Bioinformatics 2021 & K-mer     & BERT              \\
                                 & Nucleotide Transformer & BioRxiv 2023        & K-mer     & BERT              \\
                                 & DNABERT2               & ICLR 2024           & BPE       & BERT              \\
                                 & GENA-LM                & BioRxiv 2023        & BPE       & BERT              \\ \hline
\multirow{6}{*}{Convolution-based}             & Hyena-DNA              & NeurIPS 2024        & Char      & GPT               \\ 
                                &Caduceus &ICML 2024 
                                &Char &BERT     \\
                                &CNN &BMC Genomic Data 2023&One-hot&Scratch
                                \\
                                &SpliceAI &Cell 2019 & One-hot &Scratch
                                \\
                                &DeepSTARR &Nature genetics 2022 & One-hot & Scratch\\
                                &Orca& Nature genetics 2022&One-hot&Scratch\\

\bottomrule
\end{tabular}
\end{adjustbox}
    \label{support models}
    \vspace{-1.0em}
\end{table}

GenBench has developed ten key genomic foundational models within a cohesive framework, comprising four attention-based models and six convolution-based models. These models are outlined in Table~\ref{support models}, which details the associated conference/journal, the tokenizer types used, and their respective training strategies. The initial attention-based model employs a k-mer tokenizer, whereas the more recent attention-based model utilizes Byte Pair Encoding (BPE). HyenaDNA and Caduceus adopt a Char tokenizer due to their subquadratic space complexity and linear space complexity. Besides, we have included expert models for particular downstream tasks for comparison, named SpliceAI
~\cite{jaganathan2019predicting}, DeepSTARR~\cite{de2022deepstarr}, CNN~\cite{grevsova2023genomic}, and Orca~\cite{zhou2022sequence}.

This design closely resembles the conventional deep-learning-based language models \cite{devlin2018bert, brown2020language}, but with modifications to the tokenizers tailored for genomic sequences, taking into account the simpler structure of genomes compared to human language. In general, the tokenizer converts a sequence of nucleotides into tokens. Each token is then converted into a numerical vector and represented as a matrix 
$M$ through embedding. Depending on the method used to segment the nucleotide sequences, tokenizers can be divided into k-mer tokenizers and BPE tokenizers. A newer approach has also emerged, where each individual nucleotide is directly mapped, known as the `char tokenizer.'


\begin{table}[t!]
\caption{The dataset statistics for the tasks facilitated by GenBench are meticulously detailed, delineating the various types of analyses supported. Within the typological column, the term "Sequence Binary Classification" refers to the assignment of an entire input sequence to one of two exclusive categories, thereby yielding a dichotomous classification outcome. In contrast, "Sequence Multi-class Classification" encompasses a more expansive classification, where an input sequence is allocated to one among a plurality of classes, surpassing the binary distinction. Furthermore, "Token Multi-class Classification" signifies a classification that operates at the token level, providing a nuanced categorization with multiple potential outcomes for individual elements within the sequence. Lastly, "Regression" denotes predicting a continuous range of values, as opposed to classes.}
\begin{adjustbox}{width=\columnwidth,center}
    \setlength{\tabcolsep}{1.1mm}
\begin{tabular}{l|ll|cccc}
\toprule
Benchmark                     & Tasks                                  & Type                             & Training size & Testing size & Length                   \\ \hline
\multirow{16}{*}{Short Range} & Mouse Enhancers                       & Binary classification            & 968           & 242          & 500                      \\
                              & Coding vs Intergenomic                & Binary classification            & 75K         & 25K        & 500                      \\
                              & Human vs Worm                         & Binary classification            & 75K         & 25K        & 500                      \\
                              & Human Enhancers Cohn                  & Binary classification            & 20K         & 7K         & 500                      \\
                              & Human Enhancers Ensembl               & Binary classification            & 123K        & 3K         & 500                      \\
                              & Human Ensembl Regulatory              & Multi-class classification       & 231K        & 57K        & 500                      \\
                              & Human Nontata promoters               & Binary classification            & 27K         & 9K         & 500                      \\
                              & Human OCR Ensembl                     & Binary classification            & 14K        & 35K        & 500                      \\
                              & Drosophila Enhancers Prediction       & Regression                       & 402K        & 41K        & 128                      \\
                              & Human Core Promoter Detection         & Binary classification            & 95K         & 12K        & 70                       \\
                              & Human Transcription Factor Prediction & Binary classification            & 128K        & 5K         & 30                       \\
                              & Human Promoter Detection              & Binary classification            & 95K         & 12K        & 70                       \\
                              & Human Splice Site Detection           & Multi-class classification       & 36K         & 5K         & 80                       \\
                              & Mouse Transcription Factor Prediction & Binary classification            & 80K         & 10K        & 30                       \\
                              & Yeast Epigenetic Marks Prediction     & Binary classification            & 230K        & 29K        & 128                      \\
                              & Virus Covid Variant Classification    & Multi-class classification       & 73K         & 9K         & 256                      \\ \hline
\multirow{5}{*}{Long Range}   & Splice Site Prediction                & Multi-class classification & 146K        & 16K        & 15K                     \\
                              & Species Classification                & Multi-class classification       & 1K          & 500          & 80M                     \\
                              & Promoters Prediction                  & Binary classification            & 41K         & 12K        & 8K                     \\
                              & Genomic Structure Prediction          & Regression                       & 21           & 3          & 256M                     \\
                              & Bulk RNA Prediction                   & Regression                       & 23K         & 990          & 196K \\ \bottomrule
\end{tabular}
\end{adjustbox}
    \vspace{-1.0em}
    \label{tab:dataset_stats}
\end{table}

\subsection{Support Tasks}
Genbench covers local-to-global genomic tasks comprehensively. For simplicity, we have segmented the GenBench into short and long-range tasks based on a criterion of 1k length, considering that the sequence length significantly affects performance and complexity. The GenBench benchmark encompasses diverse genomic targets, such as enhancers, promoters, and splice sites, at different scales. The tasks involve binary sequence classification, multi-class sequence classification, multi-class token classification, and regression tasks. A summary of the datasets can be found in Table~\ref{tab:dataset_stats}.

\textbf{Short-Range Tasks.} 
Short-range tasks are characterized by input lengths of less than one thousand. Our analysis covers thirty-eight datasets related to short-range tasks, which include various types of tasks like sequence classification, variant classification, Epigenetic mark prediction, promoter detection, enhancer prediction, transcription factor detection, and splice site prediction~\cite{nguyen2024hyenadna,zhou2023dnabert,de2022deepstarr,ji2021dnabert}.

\textbf{Long-Range Tasks.} 
Long-range tasks are defined as tasks with input lengths longer than 1000. Achieving state-of-the-art performance on benchmarks involving long sequences, such as the Long Range Arena (LRA)~\cite{tay2020long}, is possible. However, longer context lengths also introduce higher complexity. For instance, attention-based models exhibit quadratic complexity concerning input length~\cite{vaswani2017attention}. In long-range tasks, we include site annotation~\cite{jaganathan2019predicting}, species classification~\cite{nguyen2024hyenadna}, promoter prediction~\cite{fishman2023gena}, chromatin profiling~\cite{zhou2015predicting}, and genomic structure prediction~\cite{schwessinger2020deepc}. 

\subsection{Data Description}
\emph{Mouse Enhancers Ensembl}, \emph{Coding vs Intergenomic}, \emph{Human vs Worm}, \emph{Human Enhancers Cohn}, \emph{Human Enhancers Ensembl}, \emph{Human Ensembl Regulatory}, \emph{Human Nontata promoters}, and \emph{Human OCR Ensembl} were referenced from \cite{grevsova2023genomic}. In this context, "Human" and "Mouse" signify the origin of the genetic sequences, while "Enhancers," "Regulatory," "OCR," and "promoter" describe the nature of the sequences.
A regulatory gene is responsible for controlling the expression of one or more structural genes. Enhancers are specific genomic elements that regulate gene expression without requiring close proximity to the target gene. Open chromatin regions (OCR) are parts of the genome that can be easily accessed by DNA regulatory elements. On the other hand, a promoter is a segment within a gene where specific proteins bind to initiate the gene's transcription.
The term "Ensembl" in this context refers to the data resources provided by The Ensembl project \cite{howe2021ensembl}, which is involved in genome annotation and distribution of genomic information for various vertebrate species. Furthermore, this study involves the classification of genomic sequences across different species or transcript types, such as protein-coding versus non-coding sequences.

The tasks of \emph{Human Promoter Prediction}, \emph{Human Core Promoter Detection}, \emph{Human Transcription Factor Prediction}, \emph{Human Splice Site Detection}, \emph{Mouse Transcription Factor Prediction}, \emph{Yeast Epigenetic Marks Prediction}, and \emph{Virus Covid Variant Classification} adopted from \cite{zhou2023dnabert} encompass a variety of objectives. These tasks involve predicting different region types, such as promoters, transcription factor binding sites, and splice sites across multiple animal genes, as well as predicting variants of the Covid virus based on provided gene sequences.

We include \emph{Splice Site Prediction}, \emph{Promoter Prediction}, and \emph{Drosophila Enhancer Detection} in our assessment following the methodology described in \cite{fishman2023gena}. These datasets are known for their extensive sequences and varied tasks. They comprise sequences exceeding 1000 base pairs, covering a range of tasks like token classification, sequence-level classification, and regression. In particular, \emph{Drosophila enhancers prediction} involves a two-class regression, where the goal is to predict two float values for every 249-base pair sequence, one for housekeeping and one for developmental enhancer scores. These tasks are crafted to evaluate the model's performance under diverse conditions.

Moreover, we have integrated \emph{genomic structure prediction} as presented in \cite{zhou2022sequence}. This prediction specifically examines how structural variants impact genome organization at various scales. In order to enhance practicality, we have limited the length to 6000-bp. Additionally, we have included \emph{Species Classification} from \cite{nguyen2024hyenadna}, which has heightened the complexity of classification by encompassing a larger number of species.
We incorporated the task of \emph{Bulk RNA expression} to evaluate the model's performance within a lengthy context \cite{kao2024advancing}. Bulk RNA-sequencing is a biological assay that gauges the average gene expression from a group of cells within a specific tissue. In this particular task, the input is the sequence from the human
reference genome that is centered around the Transcription Start Site (TSS). The resulting output is a unified vector consisting of continuous values that indicate the bulk RNA levels of a gene across 218 distinct tissue types.

\subsection{Evaluation Metrics}
We thoroughly assess the performance of the models supported for the tasks mentioned above by employing a range of metrics. These metrics are tailored to the specific characteristics of each task.

\vspace{-0.5em}
\begin{itemize}[leftmargin=*]
    \item \textbf{Error metrics:} We use cross-entropy to assess the variance between the anticipated outcomes and the actual targets in both binary and multi-class classification scenarios. On the other hand, Mean Squared Error (MSE) is applied in regression tasks.
    
    \item \textbf{Accuracy metrics:} We use top-1 accuracy for classification tasks and combine the evaluation metrics of computing the Area Under the Receiver Operating Characteristic Curve (AUC-ROC)~\cite{hand2001simple}.
    
    \item \textbf{Correlation metrics:} Spearman correlation coefficient (Spearman)~\cite{sedgwick2014spearman} and the Pearson correlation coefficient (Pearson)~\cite{kowalski1972effects} for regression tasks to assess the model's accuracy.

    \item \textbf{Computational metrics:} We utilize the number of parameters and the number of floating-point operations (FLOPs) to evaluate the computational complexity of the models. 
\end{itemize}




\subsection{Codebase Structure}
Current open-source genomic foundation model codebases are typically constrained by a limited number of datasets and models. In contrast, GenBench offers a versatile and expandable framework that follows the design principles outlined in HyenaDNA\cite{nguyen2024hyenadna}. GenBench stands out for its user-friendly interface, well-organized structure, and comprehensive nature, outshining the usability of other open-source genomic foundation model codebases. For details, please refer to the Appendix~\ref{app:codebase}.

%% file: 4_exps.tex
\begin{table}[b!]
\vspace{-1em}
\caption{Top-1 Pearson score for pre-trained HyenaDNA,  DNABERT2,
GENA-LM, Nucleotide Transformer, and Caduceus, and non-pre-trained model of deepstar in short-range task of drosophila enhancers prediction regarding the developmental (dev) and housekeeping	
activity (hk).}
\begin{adjustbox}{width=\columnwidth,center}
\begin{tabular}{lcccccc}
\toprule
Dataset         & HyenaDNA($\uparrow$) & DNABERT2($\uparrow$) & GENA-LM($\uparrow$) & Nucleotide Transformer($\uparrow$) &Caduceus($\uparrow$) &DeepSTARR($\uparrow$)\\  \hline
dev    & 0.470   & \underline{0.617}   & \textbf{0.624}  & 0.612  &0.443& 0.424               \\
hk & 0.552   & 0.734   & \textbf{0.740}  & \underline{0.736}   &0.530 & 0.513             \\
Mean            & 0.511   & \underline{0.678}   & \textbf{0.682}  & 0.674    &0.486 & 0.468             \\ \bottomrule
\end{tabular}
\end{adjustbox}
\label{tab:splice_site_annotation}
\vspace{-1em}
\end{table}
\begin{table}[b!]
\caption{Short-Range Tasks Top-1 accuracy (\%) for pre-trained HyenaDNA, DNABERT, DNABERT2, GENA-LM,  Nucleotide Transformer, Caduceus, and CNN. Higher value indicate better performance.}
\begin{adjustbox}{width=\columnwidth,center}
    \setlength{\tabcolsep}{0.8mm}
\begin{tabular}{lccccccc}
\toprule
Dataset                 & HyenaDNA($\uparrow$) & DNABERT($\uparrow$) & DNABERT2($\uparrow$) & GENA-LM($\uparrow$) & NT($\uparrow$)& Caduceus($\uparrow$)& CNN($\uparrow$) \\ \hline
Mouse Enhancers         & 0.7934   & 0.8099  & 0.8182   & \underline{0.8297}  & \textbf{0.8512}            &0.8163 & 0.7008   \\
Coding vs Intergenomic  & 0.9097   & 0.9364  & 0.9358   & 0.9324  & \textbf{0.9576} &\underline{0.9372}         &0.8844       \\
Human vs Worm           & 0.9624   & 0.9584  & 0.9739   & \underline{0.9698}  & \textbf{0.9751}  &0.9557         &0.9408      \\
Human Enhancers Cohn    & 0.7296   & 0.7023  & 0.7587   & \underline{0.7563}  & \textbf{0.7612}   &0.7376     &0.7080         \\
Human Enhancers Ensembl & 0.9033   & 0.8919  & 0.9075   & \underline{0.9107}  & \textbf{0.9244}    &0.8448     &0.7637    \\
Human Ensembl Regulatory        & 0.8462   & \underline{0.938}   & 0.8832   & 0.8810  & \textbf{0.9403}    &     0.7367  &0.8616      \\
Human Nontata Promoters & 0.9445   & 0.8713  & \underline{0.9524}   & \textbf{0.9660}  & 0.9295  &0.8885    &0.8564           \\
Human OCR Ensembl       & \underline{0.7914}   & 0.7496  & 0.7582   & 0.7898  & 0.8042       &\textbf{0.8176}  &0.6947        \\ 
Human Core Promoter Detection         & 0.8440   & 0.8491  & 0.8257   & 0.8140  & \textbf{0.8541}     &\underline{0.8505}  &0.8003          \\
Human Transcription Factor Prediction & 0.6976   & 0.7840  & 0.8218   & \underline{0.8240}  & \textbf{0.8262}            &0.6928  &0.6672   \\
Human Promoter Detection              & 0.7295   & 0.8393  & 0.8993   & \underline{0.9001}  & \textbf{0.9390}               &0.7322 &0.6875  \\
Human Splice Site Detection           & 0.5660    & 0.8721  & 0.8813   & \underline{0.9178}  & \textbf{0.9481}            &0.5674  &0.5666   \\
Mouse Transcription Factor Prediction & 0.6535   & 0.7393  & \underline{0.8269}   & 0.8265  & \textbf{0.8502}             &0.6519 &0.6081   \\
Yeast Epigenetic Marks Prediction     & 0.6301   & 0.7203  & \textbf{0.8022}   & 0.7829  & \underline{0.7845}             &0.6378 &0.6071   \\
Virus Covid Variant Classification    & 0.3770   & 0.5990  & \underline{0.7195}   & \textbf{0.7033}  & 0.6939            &0.3794   &0.1974  \\ \bottomrule
\end{tabular}
\end{adjustbox}
\label{short range task}
\end{table}

\vspace{-1em}
\section{Experiment and Analysis}
We performed thorough experiments on the mentioned tasks to evaluate the effectiveness of the supported methods in GenBench. The \textbf{bold} value indicates the best performance, and the \underline{underline} value indicates the second-best performance. Due to the limitation of computational resources, it is worth noting that we set the maximum length to 30k in the long sequence task, and the Nucleotide Transformer is sometimes written NT. A detailed analysis of the results is provided to better understand the genome foundation model. For implementation specifics, please refer to the Appendix~\ref{implementation details}. 

\subsection{Short Range Tasks}
An experimental study was carried out to evaluate how different computational models perform in handling short-range tasks. We draw several conclusions from the results. The details in Table~\ref{short range task}.

\begin{table}[t!]
\caption{Top-1 AUC-ROC Score for pre-trained HyenaDNA,  DNABERT2,
GENA-LM, Nucleotide Transformer, and Caduceus in long-range task of splice site prediction.}
\begin{adjustbox}{width=\columnwidth,center}
    \setlength{\tabcolsep}{1.0mm}
\begin{tabular}{lcccccc}
\toprule
Dataset         & HyenaDNA($\uparrow$) & DNABERT2($\uparrow$) & GENA-LM($\uparrow$) & Nucleotide Transformer($\uparrow$) &Caduceus($\uparrow$) &SpliceAI($\uparrow$) \\ \hline
Splice donar    & 0.574   & \underline{0.635}   & 0.629  & 0.557  &\textbf{0.642}  &0.574            \\
Splice acceptor & 0.723   & 0.707   & \underline{0.730}  & 0.722   &\textbf{0.740} &   0.691          \\
Mean            & 0.648   & 0.671   & \underline{0.679}  & 0.639    &\textbf{0.691} & 0.632             \\ \bottomrule
\end{tabular}
\end{adjustbox}
\label{tab:splice site annotation}
\vspace{-1.5em}
\end{table}

\begin{table}[t!]
\caption{Top-1 accuracy for pre-trained HyenaDNA,  DNABERT2,
GENA-LM, Nucleotide Transformer, and Caduceus in long range task of Species Classification and Promoters Prediction. Higher values indicate better performance.}
\begin{adjustbox}{width=\columnwidth,center}
    \setlength{\tabcolsep}{0.8mm}
\begin{tabular}{lccccc}
\toprule
Dataset                & HyenaDNA($\uparrow$) & DNABERT2($\uparrow$) & GENA-LM($\uparrow$) & Nucleotide Transformer($\uparrow$)&Caduceus($\uparrow$) \\ \hline
Species Classification & 0.6770    & 0.7470    & \underline{0.7730}   & \textbf{0.7860}     &0.7110             \\
Promoters Prediction   & 0.8875   & 0.9758   & \underline{0.9795}  & \textbf{0.9890}      &0.9302            \\ \bottomrule
\end{tabular}
\end{adjustbox}
\label{long range task}
\vspace{-1em}
\end{table}

\textbf{Convolution-based models consistently show lower performance compared to attention-based models, particularly on challenging tasks.} 
In binary classification scenarios, Hyena-DNA demonstrates a notable decrease in accuracy, exhibiting a reduction of 0.053 compared to the attention-based models. This discrepancy becomes even more pronounced in multi-class classification tasks, where the accuracy gap widens to 0.239. Similarly, another model, Caduceus, shows a comparable pattern, with a 0.053 accuracy gap in binary classification tasks and a substantially larger margin of 0.274 in multi-class classification assignments. When comparing these models to the performance of pre-trained models, the discrepancy becomes even more striking. The CNN model, in particular, achieved lower accuracy, with the accuracy gap reaching up to 0.387. These findings highlight the challenges faced by convolution-based models like Hyena-DNA and Caduceus, especially when compared to attention-based and pre-trained models.

\begin{figure*}[t!]
\centering
    \hspace{-1.0em}
    \subfloat{
    \includegraphics[width=0.425\textwidth]{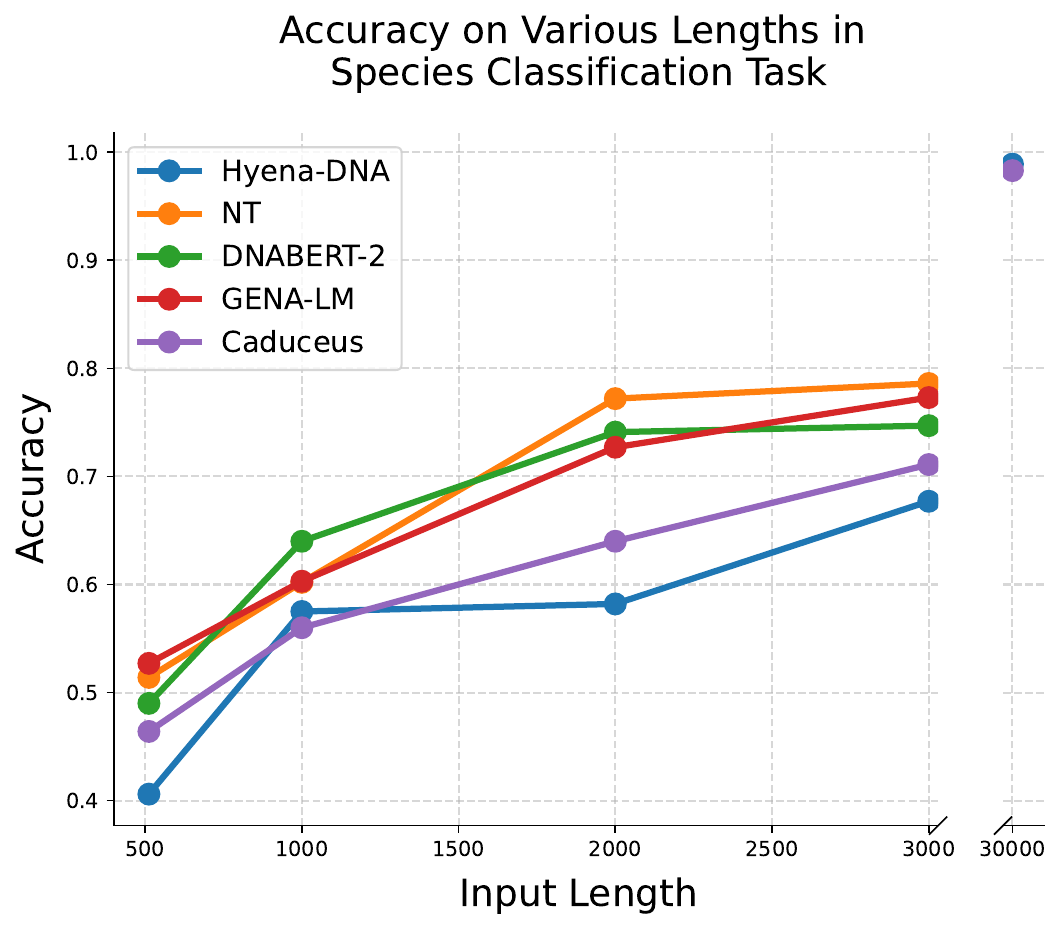}}
    \hspace{1.5em}
    \vspace{-1.0em}
    \subfloat{
    \includegraphics[width=0.425\textwidth]{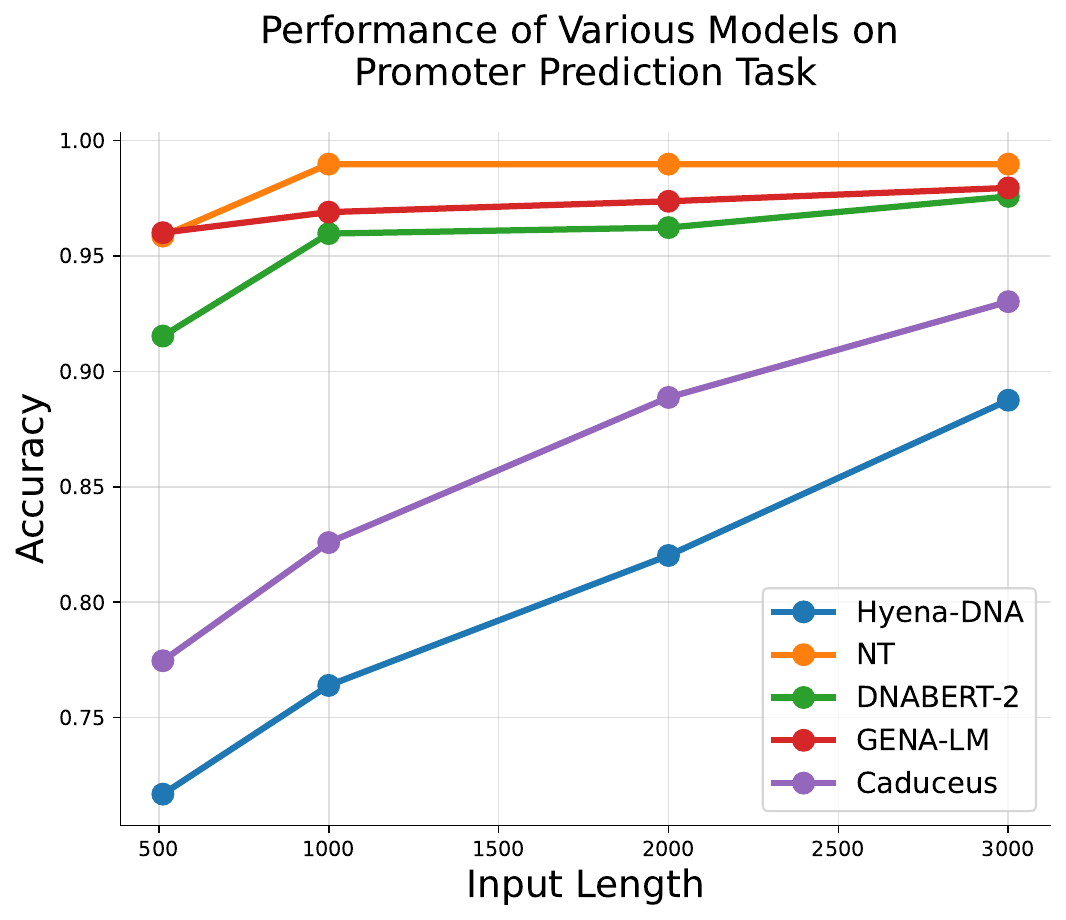}}
    \hspace{1.5em}
    \subfloat{
    \vspace{-1.0em}
    \includegraphics[width=0.875\textwidth]{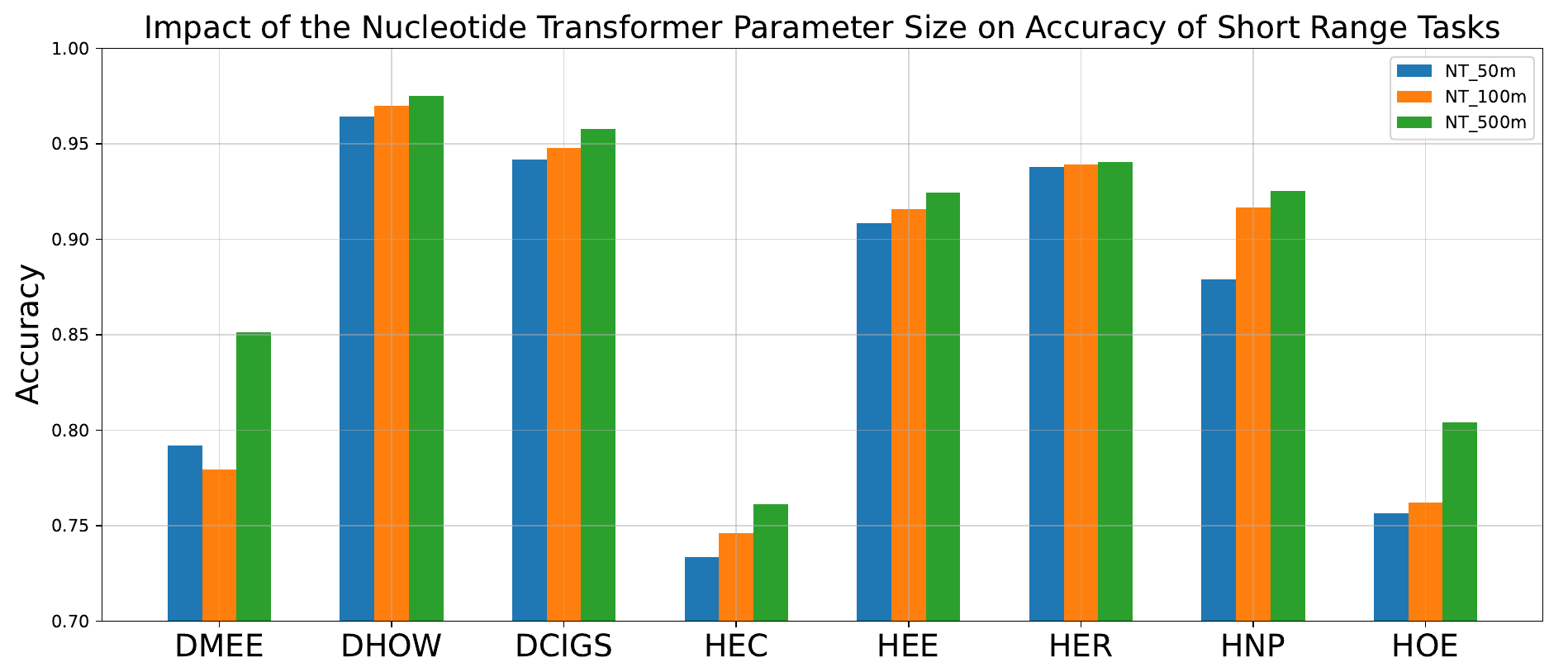}}
\vspace{-0.5em}
    \caption{Impact of length and parameter size on accuracy: (a) Accuracy variation across different lengths in species. (b) Accuracy variation across different lengths in promoter prediction tasks. (c) Evaluation of NT accuracy on short-range tasks with parameter sizes of 50M, 100M, and 500M.}
    \label{main:c}
    \vspace{-1.0em}
\end{figure*}

\textbf{The size of parameters plays a crucial role in determining performance}. 
Similar to NLP, the scaling law works in short-range tasks. 
Among the models analyzed, the Nucleotide Transformer, which boasts the largest parameter size, outperformed others on 11 out of 15 datasets. Subsequently, GENA-LM and DNABERT2, having similar parameter sizes, excelled on the 2 and 1 datasets, respectively. Noteworthy is the performance disparity exhibited by DNABERT compared to other attention-based models, with variances ranging from 0.0050 to 0.1401, despite sharing a similar architecture with Nucleotide Transformer albeit possessing the smallest parameter size. To delve deeper into the impact of parameter size, we examined the performance of Nucleotide Transformers with parameter sizes of 50, 100, and 500 million. The findings depicted at the bottom of Figure~\ref{main:c}.

\begin{table}[b!]
\vspace{-1em}
\caption{Top-1 Pearson and MSE for pre-trained Orca, HyenaDNA, DNABERT2, Caduceus, and Nucleotide Transformer in the long-range task of Genomic Structure Prediction.}
\begin{adjustbox}{width=1.0\columnwidth,center}
    \setlength{\tabcolsep}{0.6mm}
\begin{tabular}{ccccccccccccccccc}
    \toprule
\multicolumn{1}{l}{Dataset}                   & \multicolumn{2}{c}{Orca} & \multicolumn{2}{c}{HyenaDNA} &  \multicolumn{2}{c}{Caduceus}&  \multicolumn{2}{c}{DNABERT2}&  \multicolumn{2}{c}{GENA-LM}&  \multicolumn{2}{c}{NT}  \\ \hline
 & Pearson($\uparrow$)     & MSE($\downarrow$)        & Pearson($\uparrow$)       & MSE($\downarrow$)          &Pearson($\uparrow$)       & MSE($\downarrow$)&Pearson($\uparrow$)       & MSE($\downarrow$)&Pearson($\uparrow$)       & MSE($\downarrow$)&Pearson($\uparrow$)       & MSE($\downarrow$)        \\
                                              H1-ESC& \underline{0.4543}       & \underline{0.0175}    & 0.4357         & 0.0184    & \textbf{0.5024}         & \textbf{0.0168}
                                            & - & - & - & - & - & -
                                              \\ 
                                                HFF
                                              & \textbf{0.4350}       & \textbf{0.0911}    & 0.3103         & \underline{0.1013}    & \underline{0.3536}         & 0.1047       & - & - & - & - & - & -
                                              \\ \bottomrule
\end{tabular}
\end{adjustbox}
\label{Genomic Structure Prediction}
\vspace{-1.5em}
\end{table}

\begin{table}[b!]
\caption{ Top-1 Spearman and MSE for pre-trained HyenaDNA, DNABERT2, and Caduceus in long-range task of Bulk RNA Expression.}
\begin{adjustbox}{width=\columnwidth,center}
    \setlength{\tabcolsep}{0.6mm}
\begin{tabular}{ccccccccccc}
\toprule
\multicolumn{1}{l}{Dataset}                   &  \multicolumn{2}{c}{HyenaDNA} &  \multicolumn{2}{c}{Caduceus}&\multicolumn{2}{c}{DNABERT2}&\multicolumn{2}{c}{Nucleotide Transformer} &\multicolumn{2}{c}{GENA-LM}\\ \hline
 & Spearman($\uparrow$)     & MSE($\downarrow$)        & Spearman($\uparrow$)       & MSE($\downarrow$)          &Spearman($\uparrow$)       & MSE($\downarrow$)  &
 Spearman($\uparrow$)       &MSE($\downarrow$)&Spearman($\uparrow$) &MSE($\downarrow$)\\
                                              Bulk& 0.755       & 0.453    & 0.746         & 0.510    & \underline{0.767}         & \underline{0.451}       
                                              & \textbf{0.783}&\textbf{0.396}
                                              &0.748&0.496
                                              \\
                                                \bottomrule
\end{tabular}
\end{adjustbox}
\label{Genomic Structure Prediction}
\end{table}

\subsection{Long range tasks}
\textbf{As the sequence grows, the convolution-based models become more efficient.}
The long-range tasks are detailed in Table \ref{tab:splice_site_annotation}, \ref{long range task}, and \ref{Genomic Structure Prediction}. In the long-range task, the difference between attention-based and convolution-based models has been significantly reduced. For Species Classification and Promoters Prediction, DNABERT2, GENA-LM, and Nucleotide Transformer exhibit comparable performance, while HyenaDNA performs notably worse than the rest. However, in the task of splice site annotation, HyenaDNA achieves the highest performance, with no significant performance gap between them. Additionally, the computational overhead of the convolutional model is much smaller than that of the attention-based model for the same sequence length, as shown in Figure~\ref{flops}.

\textbf{Transformers collapse on Genome Structure Prediction.}
In Genomic Structure prediction, surprisingly, attention-based models fail to converge in this ``longest task'' and utilizing a pre-trained model as a backbone does not show significant benefits, as Orca achieved the second-best Pearson and MSE in H1-ESC and the best Pearson and MSE in HFF, respectively. We can find a greater resemblance to image data when nucleotide sequences become longer: short vocabulary (4 bases ATCG and 256 RGB values) and fixed data rules (chromosome and image content). This also explains the better performance of the convolutional networks in long sequence tasks.

\subsection{Effect of Length}
We conducted an extensive analysis to evaluate the impact of sequence length on model performance in long-range genomic tasks. Specifically, we used input sequences of varying lengths—512, 1000, 2000, and 3000 base pairs (bp)—to assess the performance of four models: Hyena-DNA, Nucleotide Transformer, DNABERT-2, and GENA-LM in both species and promoter prediction tasks. Additionally, to explore the potential of convolution-based models, we tested a significantly longer input sequence of 30,000 bp with Hyena-DNA, focusing exclusively on the species prediction task. The results of these analyses are depicted in the top right and left sections of Figure~\ref{main:c}. From the data, it is evident that increasing the sequence length consistently enhances performance across all the models tested. This trend is particularly pronounced with Hyena-DNA, which, despite trailing behind attention-based models at shorter context lengths, exhibits superior performance with longer contexts. This improvement underscores the advantages of using extended context lengths in genomic sequence analysis. However, this benefit is not without its challenges. In tasks like promoter prediction, where input length is inherently capped at 3000 bp, Hyena-DNA’s reliance on longer sequences becomes a limiting factor. This limitation presents a significant area for future research, aiming to optimize model performance within these constraints and potentially develop novel approaches to leverage longer sequences more effectively within the confines of specific genomic datasets.

\begin{figure*}[t!]
    \centering
    \includegraphics[width=\textwidth]{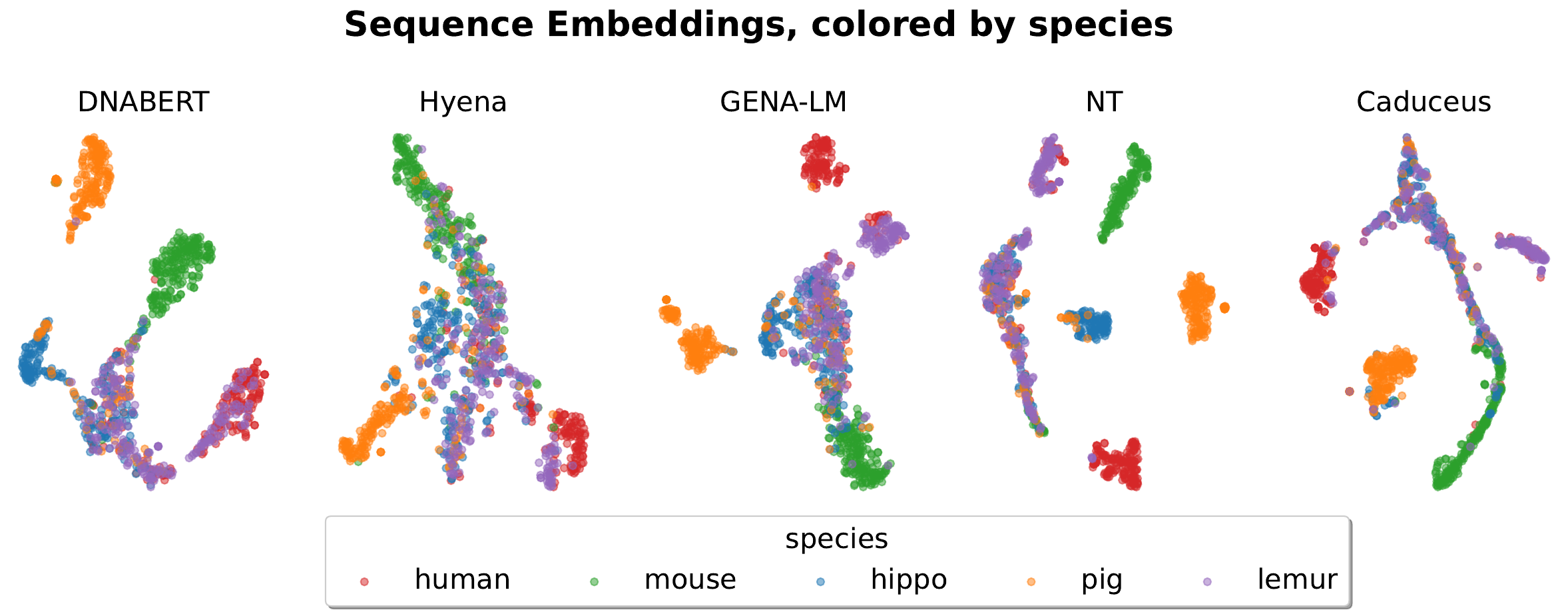}
    \caption{The t-SNE visualization of DNA embedding for foundation model in species classification. Including embedding for DNABERT2 with an accuracy of 0.742, embedding for HyenaDNA with an accuracy of 0.655, embedding for GENA-LM with an accuracy of 0.747, embedding for the NT with an accuracy of 0.761, and embedding for Caduceus with an accuracy of 0.703.}
    \label{tsne}
    \vspace{-1em}
\end{figure*}

\subsection{Gene Clustering}
In this section, we examine the fine-tuned embedding models HyenaDNA, DNABERT2, GENA-LM, Caduceus, and Nucleotide Transformer. These models are utilized to encode gene sequences from various species. To visualize the embeddings, we extract the representations from the final hidden layer of each model and apply t-distributed Stochastic Neighbor Embedding (t-SNE)~\cite{van2008visualizing}. The visualization, presented in Figure \ref{tsne}, reveals clear clusters that offer both visual and quantitative insights. For instance, the Nucleotide Transformer (NT), which demonstrates the highest accuracy among the models, shows well-separated embeddings for distinct species, indicating effective differentiation. In contrast, HyenaDNA, which has the lowest accuracy, displays less differentiation among the embeddings of different species, suggesting that its representations are less distinct. This visualization underscores the varying capabilities of distinguishing between gene sequences from different species, with NT excelling in accuracy and clarity of separation, while HyenaDNA struggles in comparison. From the results, it is clear that the k-mer based approaches have a more significant advantage.

\subsection{Computational Cost}
\begin{wrapfigure}{r}{0.4\textwidth}
    \vspace{-2em}
    \centering
    \includegraphics[width=0.95\linewidth]{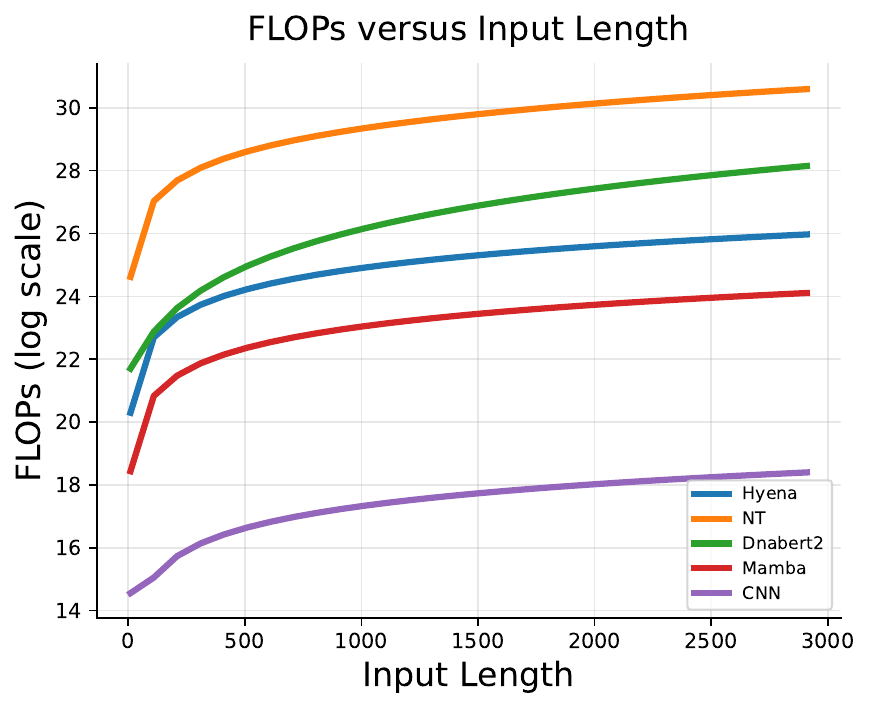}
    \vspace{-0.5em}
    \caption{Flops versus input length}
    \label{flops}
    \vspace{-2em}
\end{wrapfigure}
Being able to handle long sequences is a critical step in GFM. Therefore, we compared the floating-point operations per second (FLOPS) as a metric to evaluate the computational efficiency of each model relative to the various input lengths, as shown in Figure~\ref{flops}. Typically, attention-based models demonstrate significantly higher computational capabilities, followed by attention-free foundational models. Simple CNN models, on the other hand, exhibit the lowest computational cost. Regarding computational efficiency, it is worth considering how to combine the performance of Attentions with the speed of Convolutions.

%% file: 5_conclusions.tex
\section{Conclusion and Dicussion}
\label{sec:conclusion}
This paper presents GenBench, a comprehensive benchmark for Genomic foundational models featuring ten representative models covering a broad spectrum of challenging tasks from local to global view of genomics. GenBench classifies existing approaches into attention-based and convolution-based GFMs. 
A general Genomic foundational model architecture is proposed to enhance understanding of genomic foundational models. 
Extensive experiments are carried out to systematically assess the performance of the models supported across various tasks. In short-range tasks, attention-based models excel at capturing intrinsic information, while attention-free models achieve comparable yet less efficient performance. In long-range tasks, the performance difference between attention-based and convolution-based models becomes narrower. Furthermore, increasing input length can significantly enhance performance, particularly in extending gene context.

\textbf{Limitations.} 
Despite the multifaceted comparisons of GFMs, GenBench is basically stuck on downstream task prediction, and comparisons on pre-training are lacking. For example, the impact of pre-training data under different model structures, \textit{etc.} In addition, we have not verified the performance of GFM on whole chromosome due to the limitation of computational resources. And finally, exploring generative tasks in genomics is also interesting and worth considering in the future.

\textbf{Insights.} 
This work intends to offer valuable insights and serve as a point of reference for future research endeavors.
\textbf{Insight for model design:}
a hybrid structure of attention and convolutional networks is a feasible and efficient solution. Due to the property that the longer the sequence length, the more stable and better the model performance is, the focus is on the use of convolution supplemented by attention.
\textbf{Insight for training strategy:} From the experimental results, the combined approach of K-mer and BERT for training is more suitable for genomic data, whose embedding implies a richer biological significance.



%% file: 7_checklist.tex
\clearpage
\section*{Checklist}

The checklist follows the references.  Please
read the checklist guidelines carefully for information on how to answer these
questions.  For each question, change the default \answerTODO{} to \answerYes{},
\answerNo{}, or \answerNA{}.  You are strongly encouraged to include a {\bf
justification to your answer}, either by referencing the appropriate section of
your paper or providing a brief inline description.  For example:
\begin{itemize}
  \item Did you include the license to the code and datasets? \answerYes{See Section~\ref{app:codebase}.}
  \item Did you include the license to the code and datasets? \answerNo{The code and the data are proprietary.}
  \item Did you include the license to the code and datasets? \answerNA{}
\end{itemize}
Please do not modify the questions and only use the provided macros for your
answers.  Note that the Checklist section does not count towards the page
limit.  In your paper, please delete this instructions block and only keep the
Checklist section heading above along with the questions/answers below.

\begin{enumerate}

\item For all authors...
\begin{enumerate}
  \item Do the main claims made in the abstract and introduction accurately reflect the paper's contributions and scope?
    \answerYes{}
  \item Did you describe the limitations of your work?
    \answerYes{See Section \ref{sec:conclusion}.}
  \item Did you discuss any potential negative societal impacts of your work?
    \answerYes{}
  \item Have you read the ethics review guidelines and ensured that your paper conforms to them?
    \answerYes{}
\end{enumerate}

\item If you are including theoretical results...
\begin{enumerate}
  \item Did you state the full set of assumptions of all theoretical results?
    \answerNA{}
  \item Did you include complete proofs of all theoretical results?
    \answerNA{}
\end{enumerate}

\item If you ran experiments (e.g. for benchmarks)...
\begin{enumerate}
  \item Did you include the code, data, and instructions needed to reproduce the main experimental results (either in the supplemental material or as a URL)?
    \answerYes{}
  \item Did you specify all the training details (e.g., data splits, hyperparameters, how they were chosen)?
    \answerYes{}
  \item Did you report error bars (e.g., with respect to the random seed after running experiments multiple times)?
    \answerYes{}
\item Did you include the total amount of compute and the type of resources used (e.g., type of GPUs, internal cluster, or cloud provider)?
    \answerYes{}
\end{enumerate}

\item If you are using existing assets (e.g., code, data, models) or curating/releasing new assets...
\begin{enumerate}
  \item If your work uses existing assets, did you cite the creators?
    \answerYes{}
  \item Did you mention the license of the assets?
    \answerNo{The used code and data are open-source and under the MIT license for research usage.}
  \item Did you include any new assets either in the supplemental material or as a URL?
    \answerYes{}
  \item Did you discuss whether and how consent was obtained from people whose data you're using/curating?
    \answerYes{}
  \item Did you discuss whether the data you are using/curating contains personally identifiable information or offensive content?
    \answerNA{The used data has undergone ethical review.}
\end{enumerate}

\item If you used crowdsourcing or conducted research with human subjects...
\begin{enumerate}
  \item Did you include the full text of instructions given to participants and screenshots, if applicable?
    \answerNA{}
  \item Did you describe any potential participant risks, with links to Institutional Review Board (IRB) approvals, if applicable?
    \answerNA{}
  \item Did you include the estimated hourly wage paid to participants and the total amount spent on participant compensation?
    \answerNA{}
\end{enumerate}

\end{enumerate}

%% file: 6_appendix.tex
\appendix
\section{Codebase Overview}
\label{app:codebase}
\begin{figure}[b!]
  \centering  \includegraphics[width=1.5\linewidth]{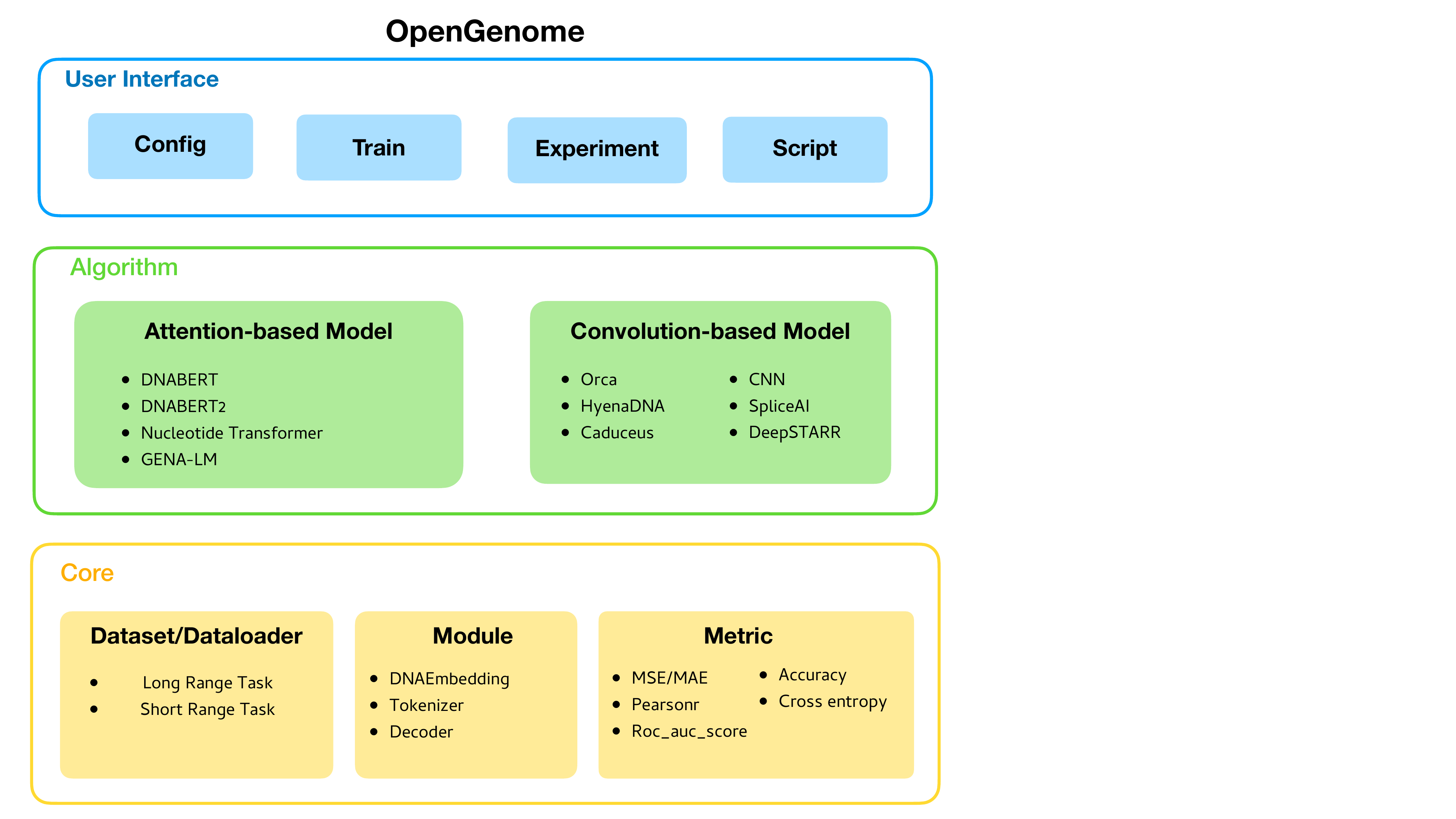}
  \caption{The graphical overview of GenBench.}
  \label{codebase structure}
\end{figure}

In this section, we present a comprehensive overview of the codebase structure of GenBench. The
codebase is organized into three abstracted layers, namely the core layer, algorithm layer, and user interface layer, arranged from the bottom to the top, as illustrated in Figure~\ref{codebase structure}. Our codebase is under Apache-2.0 license, like HyenaDNA~\cite{nguyen2024hyenadna}.

\textbf{Core Layer} The core layer of GenBench includes key elements like data loaders for supported datasets, fundamental modules for supported models, and metrics for evaluation. Data loaders provide a standardized way of loading and preprocessing data. The modules contain essential unit implementations of supported models. Metrics offer a consistent method for evaluating results. This core layer sets the groundwork for upper layers to ensure adaptable usage.

\textbf{Algorithm Layer} The algorithm layer encompasses the implementations of the supported models, which are divided into two main categories: attention-based and convolution-based models. These implementations are developed using the PyTorch framework and closely adhere to the methodologies described in the original research papers and their official open-source code. To enhance convenience, we directly incorporate the pretrained model from Huggingface. The algorithm layer ensures the compatibility, reliability, and reproducibility of the supported algorithms by abstracting common elements and preventing code duplication, thus facilitating easy and flexible integration of customized algorithms. Moreover, this layer provides a standardized interface that simplifies tasks such as model training, evaluation, and testing. By offering a consistent interface, the algorithm layer enhances user-friendliness and promotes seamless experimentation with the models.

\textbf{User Interface Layer} The user interface layer includes configurations, training, Experiments, and scripts to support the basic functions of GenBench. It provides user-friendly tools for creating visualizations. This layer is designed to be intuitive, allowing users to easily train, evaluate, and test the algorithms it supports. Through detailed parameter settings in the configurations, the user interface layer offers a unified interface that enables users to replicate the results presented in this paper without the need for extra steps.

\section{Implementation Details}
\label{implementation details}
The table presented in Table \ref{hyperparameter} outlines the hyperparameters utilized in the various models supported across different datasets. Each model's hyperparameters consist of layers, a width of the hidden dimension, parameter size, learning rate, embed dropout, residual dropout, optimizer, optimizer momentum, training epochs, batch size, LR scheduler, and reverse complement augmentation. It is important to note that sequence length is task-dependent and not directly related to the model.

\subsection{Model Description}
\textbf{DNABERT} \cite{ji2021dnabert} represents the pioneering deep learning approach that incorporates the concept of Bidirectional Encoder Representations from Transformers (BERT) model \cite{devlin2018bert} within the context of genomic DNA. Similar to BERT, DNABERT follows a pre-training—fine-tuning framework. In the pretraining phase, a portion of k contiguous tokens, covering 15\% of the sequence, is randomly masked, prompting DNABERT to forecast the masked sequences based on the remaining context. The training dataset is derived from the human genome using a direct non-overlapping splitting and random sampling approach, with sequence lengths ranging from 5 to 510.

\textbf{Nucleotide Transformer} utilizes an encoder-only transformer architecture. The models are trained using the BERT methodology. The Nucleotide Transformer employs three distinct datasets for pre-training the model: The Human reference genome dataset, The 1000G dataset, and The Multispecies dataset \cite{dalla2023nucleotide}.

\textbf{DNABERT-2} utilizes the Transformer Encoder architecture, providing flexibility in input length and enhanced computational and memory efficiency. It replaces learned positional embeddings with Attention with Linear Biases (ALiBi)\cite{press2021train} and incorporates FlashAttention \cite{dao2022flashattention} and Low Precision Layer Normalization. The model is pretrained on The Human Genome dataset and The Multi-Species Genome dataset \cite{zhou2023dnabert}.

\textbf{GENA-LM} model utilizes the Transformer Encoder architecture and has been trained on the Human T2T v2 genome assembly dataset.

\textbf{Hyena-DNA} utilizes a decoder-only design, composed of a series of blocks containing a Hyena operator. It is pretrained using the human reference genome.

\textbf{Caduceus} is a group of bidirectional long-range DNA sequence models that are the pioneers in supporting RC equivariant language modeling. Caduces employ pre-training and fine-tuning techniques with MambaDNA as their foundation.

The convolution-based deep learning models such as \textbf{CNN}, \textbf{SpliceAI}, \textbf{DeepSTARR}, and \textbf{Orca} are specifically developed to predict distinct genomic features. These models are trained from scratch using specialized datasets instead of being pretrained on general genomic sequences

\begin{table}[]
\caption{Hyperparameter ranges used to fine-tune all models for all datasets.}
\begin{adjustbox}{width=\columnwidth,center}
\begin{tabular}{lcccccc}
\toprule
Configuration               & HyenaDNA                                     & DNABERT                      & DNABERT2                     & GENA-LM                                    & Nucleotide Transformer  &    Caduceus                                     \\ \hline
Layers                      & 8                                            & 12                           & 12                           & 12                                         & 29  &16                                                             \\
Width                       & 256                                          & 768                          & 768                          & 768                                       & 1024     &256                                                        \\
Parameters                  & 6.6 M                                        & 86.1M                        & 117M                         & 113M                                       & 498M   &7.9M                                                          \\
Optimizer                   & \multicolumn{6}{c}{AdamW}                                                                                                                                                                                                  \\
Optimizer momentum          & \multicolumn{6}{c}{$\beta_1$, $\beta_2$ = 0.9, 0.999}                                                                                                                                                                                    \\
Training epoch              & \multicolumn{6}{c}{100}                                                                                                                                                                                                    \\
Batch size                  & \multicolumn{6}{c}{128-256}                                                                                                                                                                                                \\
Learning rate               & 1e-4 to 6e-4                                 & 3e-5                         & 3e-5                         & 5e-5                                       & 1e-5   &1e-4 to 1e-3                                                          \\
LR scheduler                & \multicolumn{5}{c}{Cosine decay}                                                                                                                                                                                           \\
Weight decay (model)        & 0.1                                          & 0.1                          & 0.1                          & 0.1                                        & 0.1     &0.1                                                         \\
Weight decay (Hyena layers) & 0                                            & \textbackslash{}             & \textbackslash{}             & \textbackslash{}                           & \textbackslash{}  &\textbackslash{}                                               \\
Embed dropout               & 0.1                                          & 0.1                          & 0.1                          & 0.1                                        & 0.0           &0.0                                                   \\
Resid dropout               & 0.0                                          & 0.0                          & 0.0                          & 0.0                                        & 0.0   &0.0                                                           \\
Reverse complement aug.     & \multicolumn{6}{c}{False}                                                                                                                                                                                                  \\
Sequence lengths            & 30 to 30k                                   & 30 to 512                    & 30 to 3k                     & 30 to 3k                                   & 30 to 3k    &30 to 30k                                                     \\ \bottomrule
\end{tabular}
\end{adjustbox}
\label{hyperparameter}
\end{table}

\subsection{Module Description}
\textbf{Attention} is the scaled dot product operation used to represent the relationships within the input or output sequence. This attention mechanism plays a crucial role in the Transformer model, which has been a significant advancement in deep learning \cite{devlin2018bert,radford2018improving}. The formulation of attention is as follows:
\begin{equation}
\operatorname{Attention}(Q, K, V)=\operatorname{softmax}\left(\frac{Q K^T}{\sqrt{d_k}}\right) V
\end{equation}
Where $Q$, $K$, and $V$ are mapped from the input with linear layer.

\textbf{Hyena} a class of data-controlled operators that involve a combination of multiplicative gating interactions and long convolutions, introduced by \cite{poli2023hyena}. The formulation of attention is as follows:
\begin{equation}
y=\mathrm{H}(u) v=\mathrm{D}_x^N \mathrm{~S}_h^N \cdots \mathrm{D}_x^2 \mathrm{~S}_h^2 \mathrm{D}_x^1 \mathrm{~S}_h^1 v
\end{equation}

Where $\mathrm{D}_x^n=\operatorname{diag}\left(x^n\right) \in \mathbb{R}^{L \times L}$ and $\mathrm{S}_x^n$ are Toeplitz matrix corresponding to $\mathrm{h}^n$ \cite{farenick2021operator}.

\textbf{State Space Model} is an class of sequence models have proven to be effective at handling
long-range models \cite{gu2023mamba}. The formulation of attention is as follows:
\begin{equation}
    \dot{\boldsymbol{h}}(t)=\boldsymbol{A} h(t)+\boldsymbol{B} x(t), \quad y(t)=\boldsymbol{C} h(t)+\boldsymbol{D} x(t)
\end{equation}

Where $\boldsymbol{A} \in \mathbb{R}^{N \times N}$, $\boldsymbol{B} \in \mathbb{R}^{N \times 1}$, $\boldsymbol{C} \in \mathbb{R}^{1 \times N}$, and $\boldsymbol{D} \in \mathbb{R}$ are the parameters of the system.

\section{Detailed Experimental Results}
\label{detail experiment}
Table \ref{detail of GUE} presents the outcomes for tasks involving multiple datasets. Generally, the results correspond to the average performance, with few exceptions. For instance, in the Yeast Epigenetic Marks Prediction task, GENA-LM secured the top rank. However, within the H4 dataset for Yeast Epigenetic Marks Prediction, Nucleotide Transformer exhibited superior performance, indicating variability among datasets.

\begin{table}[]
\caption{The Top-1 accuracy (\%) for short-range tasks includes various datasets for pretrained models such as HyenaDNA, DNABERT, DNABERT2, GENA-LM, and Nucleotide Transformer. A higher value signifies better performance, with \textbf{bold} indicating the best performance and \underline{underline} indicating the second best performance.}
\begin{adjustbox}{width=\columnwidth,center}
\begin{tabular}{llccccccc}
\toprule
Task                                                   & Dataset       & HyenaDNA($\uparrow$) & DNABERT($\uparrow$) & DNABERT2($\uparrow$) & GENA-LM($\uparrow$) & Nucleotide Transformer($\uparrow$)&Caduceus($\uparrow$)&CNN($\uparrow$) \\ \hline
\multirow{3}{*}{Human Core Promoter Detection}         & all           & 0.8306   & \underline{0.8429}  & 0.8236   & 0.8157  & \textbf{0.8431}  & 0.8341 &0.7883             \\
                                                       & notata        & 0.8348   & 0.8377  & \underline{0.8437}   & 0.8357  & \textbf{0.8510} &0.8427  &0.7994               \\
                                                       & tata          & 0.8667   & 0.8667  & 0.8098   & 0.7906  & \underline{0.8683} &\textbf{0.8747}&0.8133                \\ \hline
\multirow{5}{*}{Human Transcription Factor Prediction} & 0             & 0.7220   & 0.8100  & \textbf{0.8380}   & \underline{0.8320}  & 0.8250 &0.7140 &0.6940               \\
                                                       & 1             & 0.7400   & 0.8390  & \underline{0.8520}   & 0.8510 & \textbf{0.8590} &0.7590 &0.7220              \\
                                                       & 2             & 0.6690   & 0.7700  & 0.8020   & \textbf{0.8120}  & \underline{0.8070}  &  0.6650&0.6310             \\
                                                       & 3             & 0.6430   & 0.6800  & 0.7550   & \textbf{0.7770}  & \underline{0.7690}  &0.6370  &0.6080            \\
                                                       & 4             & 0.7140   & 0.8210  & \underline{0.8620}   & 0.8550 & \textbf{0.8710}   &0.6890 &0.6810             \\ \hline
\multirow{3}{*}{Human Promoter Detection}              & all           & 0.7748   & 0.8767  & 0.9255   & \underline{0.9287}  & \textbf{0.9628} &0.7833 &0.7297               \\
                                                       & notata        & 0.8102   & 0.9211  & \underline{0.9661}   & 0.9616  & \textbf{0.9778}     &0.8279  &0.7468          \\
                                                       & tata          & 0.6036   & 0.7203  & 0.8065   & \underline{0.8101}  & \textbf{0.8764}     &0.5854  &0.5860         \\ \hline
Human Splice Site Detection                            & Reconstructed & 0.5660    & 0.8721  & 0.8813   & \underline{0.9178}  & \textbf{0.9481}  &0.5674  &0.5666             \\ \hline
\multirow{5}{*}{Mouse Transcription Factor Prediction} & 0             & 0.6160   & 0.6704  & 0.7444   & \underline{0.7574}  & \textbf{0.8185} &0.6012 &0.5468               \\
                                                       & 1             & 0.7275   & 0.8765  & \underline{0.9186}   & 0.9078  & \textbf{0.9306} &0.7386 &0.7018               \\
                                                       & 2             & 0.7061   & 0.8273  & \underline{0.897}    & \textbf{0.9055}  & 0.8879 &0.7485&0.6585
                                                       
                                        \\
                                                       & 3             & 0.6208   & 0.6542  & \underline{0.8417}   & 0.8292  & \textbf{0.8667}   &0.5792  &0.5667            \\
                                                       & 4             & 0.5973   & 0.6684  & 0.7332   & \underline{0.7330}  & \textbf{0.7475}  &0.5920  &0.5669             \\ \hline
\multirow{10}{*}{Yeast Epigenetic Marks Prediction}    & H3            & 0.6853   & 0.8253  & \textbf{0.8967}   & \underline{0.8893}  & 0.8820  &0.7147 &0.6544               \\
                                                       & H3K14ac       & 0.6227   & 0.6853  & \textbf{0.7791}   & 0.7603  & \underline{0.7707}   &0.6457&0.5964              \\
                                                       & H3K36me3      & 0.6332   & 0.7246  & \underline{0.7968}   & 0.7761  & \textbf{0.810} &0.6527&0.6150                \\
                                                       & H3K4me1       & 0.5940   & 0.6836  & \underline{0.7259}   & 0.7210  & \textbf{0.7700}    &0.6032&0.5808              \\
                                                       & H3K4me2       & 0.6094   & 0.6664  & \textbf{0.7339}   & \underline{0.7070}  & 0.6762   &0.5938&0.5924              \\
                                                       & H3K4me3       & 0.5690   & 0.6101  & \textbf{0.7603}   & \underline{0.7144}  & 0.7120   &0.5696&0.5709              \\
                                                       & H3K79me3      & 0.6588   & 0.7771  & \textbf{0.8406}   & \underline{0.8318}  & 0.8121      &0.6586&0.6231           \\
                                                       & H3K9ac        & 0.6360   & 0.7133  & \textbf{0.8144}   & \underline{0.7842}  & 0.7737    &0.6392&0.5989             \\
                                                       & H4            & 0.6901   & 0.8546  & \underline{0.9058}   & 0.8921  & \textbf{0.9058}    &0.7058&0.6598             \\
                                                       & H4ac          & 0.6029   & 0.6636  & \textbf{0.7686}   & \underline{0.7529}  & 0.7326    &0.5947&0.5800             \\ \hline
Virus Covid Variant Classification                     & Covid         & 0.3770   & 0.5990  & \textbf{0.7195}   & \underline{0.7033}  & 0.6939  &0.3794 &0.1974              \\ \bottomrule
\end{tabular}
\end{adjustbox}
\label{detail of GUE}
\end{table}

\section{Visulization Result}
\subsection{Genomic Structure Prediction} 
\begin{figure*}[t!]
    \centering
    
    \includegraphics[width=0.85\linewidth]{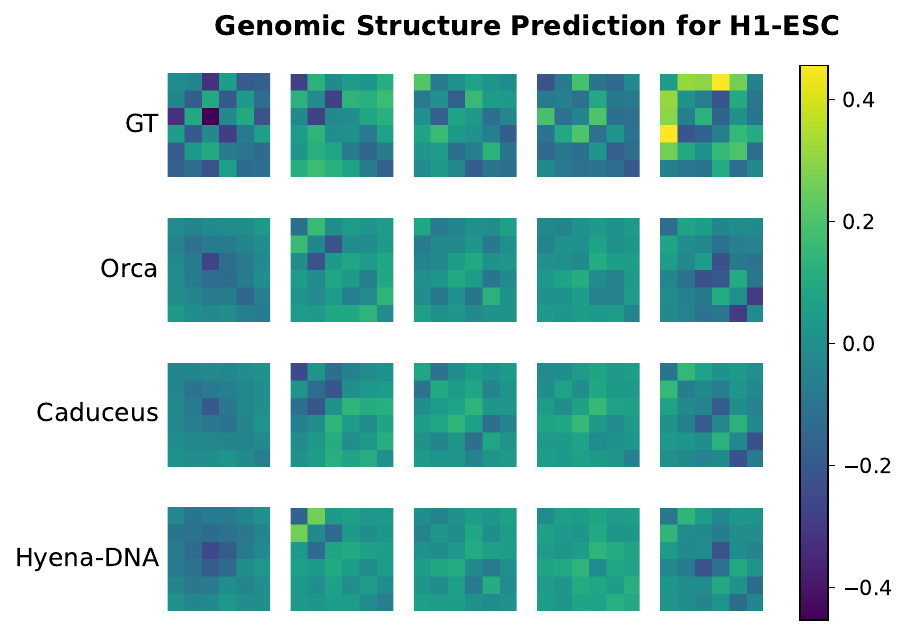}
    
    \caption{Genomic structure predictions for Orca, HyenaDNA, and Caduceus in H1-ESC visualization.}
    \label{genomic sturcture hiesc}
\end{figure*}

\begin{figure*}[t!]
    \centering
    
    \includegraphics[width=0.85\linewidth]{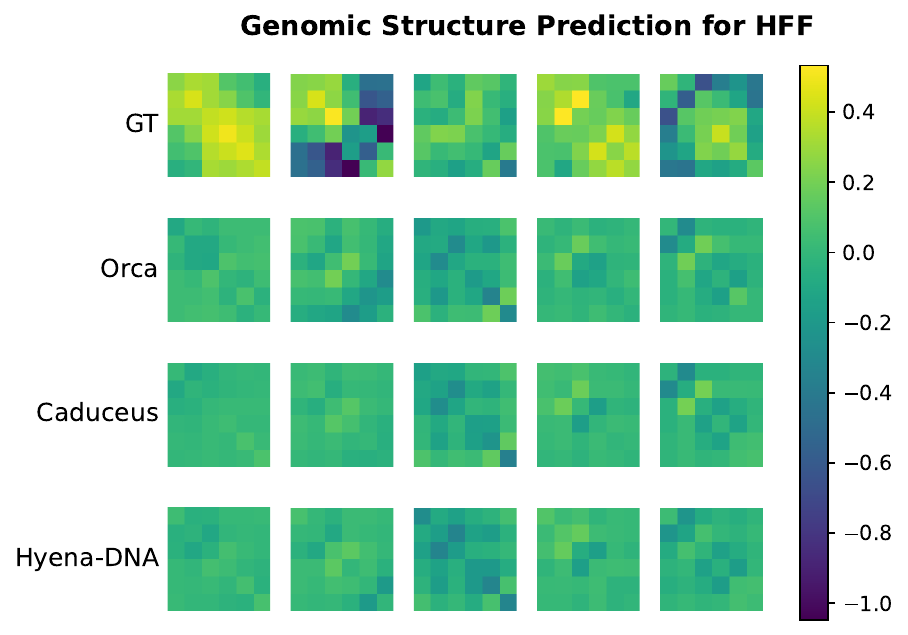}
    
    \caption{Genomic structure predictions for Orca, HyenaDNA, and Caduceus in Hff visualization.}
    \label{genomic sturcture hff}
\end{figure*}

In addition to the quantitative results presented in the main text, we also offer a visual representation for qualitative evaluation, as depicted in Figure \ref{genomic sturcture hiesc} and Figure \ref{genomic sturcture hff}. Across all models examined, we illustrate both the accurate and inaccurate prediction outcomes for comparison. It is noted that while HyenaDNA and DNABERT2 exhibit diverse predictions, Orca's predictions are relatively consistent.

\begin{figure*}[t!]
    \centering
    
    \includegraphics[width=\linewidth]{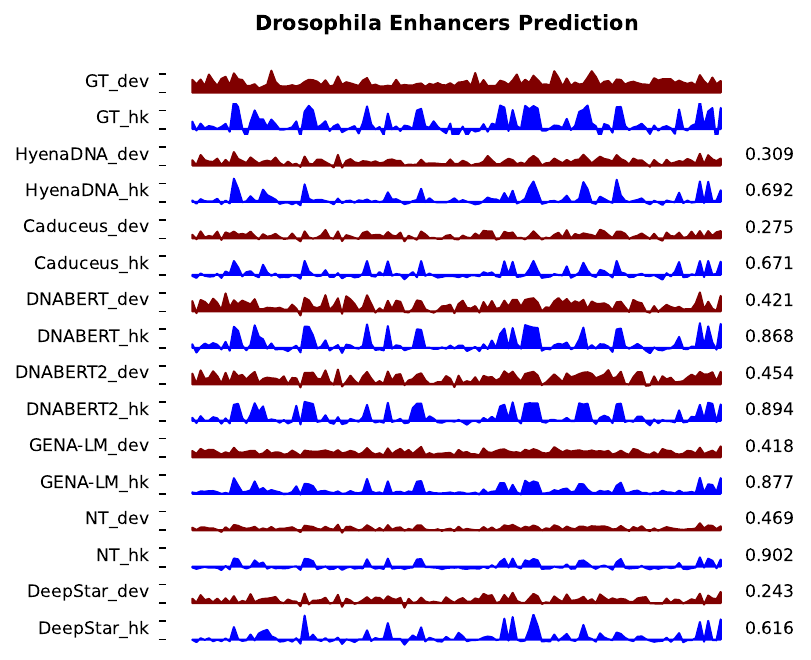}
    
    \caption{We present the visualization of Drosophila Enhancer Detection for HyenaDNA, Caduceus, DNABERT, GENA-LM, NT, and DNABERT2. The visualization illustrates the scores for both housekeeping and developmental enhancers with batch size of 128. Additionally, we include the calculation of Pearson correlation coefficient with the actual data on the right side.}
    \label{Drosophila Enhancer Detection}
\end{figure*}

\begin{figure*}[t!]
    \centering
    
    \includegraphics[width=\linewidth]{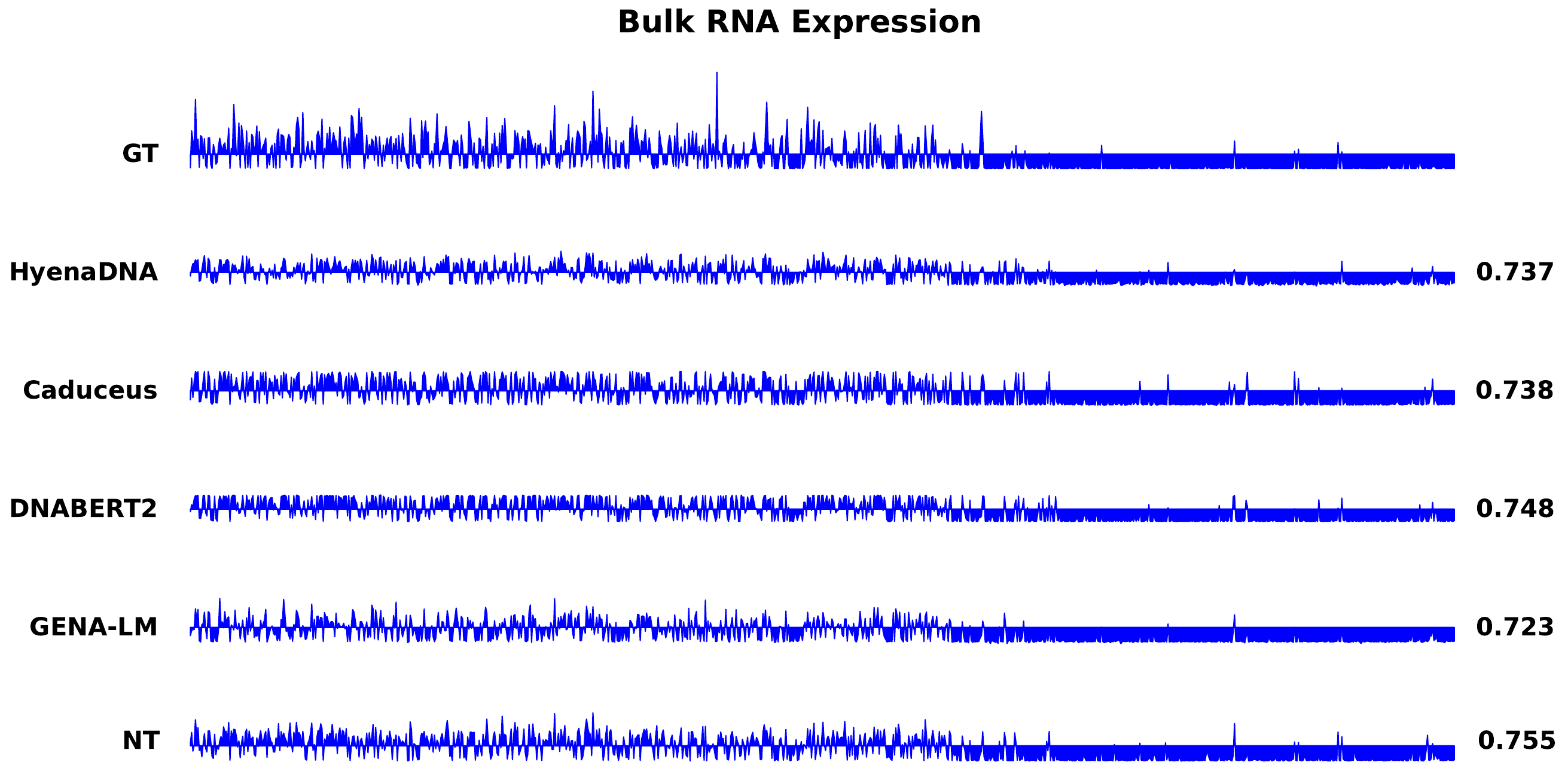}
    
    \caption{We present the visualization of Bulk RNA Expression for HyenaDNA, Caduceus, DNABERT, GENA-LM, NT, and DNABERT2. The visualization illustrates the expression levels in tissue type 0. Additionally, we include the calculation of Spearman correlation coefficient with the actual data on the right side.}
    \label{Bulk RNA expression}
\end{figure*}